\documentclass[runningheads]{llncs}
\usepackage{graphicx}
\usepackage{multirow}
\usepackage{array}
\usepackage{hhline}
\usepackage{makecell}
\usepackage{floatpag}
\usepackage{booktabs}

\begin{document}
\title{Spontaneous preterm birth prediction using convolutional neural networks}
%
%
\author{Tomasz W\l odarczyk\inst{1, 4}, Szymon P\l otka\inst{1, 4}, Przemys\l aw Rokita\inst{1}, Nicole Sochacki-W\'ojcicka\inst{2, 3}, Jakub W\'ojcicki\inst{2, 3}, Micha\l\space Lipa\inst{2, 3}, Tomasz Trzci\'nski\inst{1, 5}}
\authorrunning{T. W\l odarczyk et al.}

\institute{Warsaw University of Technology, Warsaw, Poland \and Medical University of Warsaw, Warsaw, Poland \and Ernest W\'ojcicki Prenatal Medicine Foundation, Warsaw, Poland \and Fetai Health Ltd. \and Tooploox}
\maketitle              
\begin{abstract}
An estimated 15 million babies are born too early every year. Approximately 1 million children die each year due to complications of preterm birth (PTB). Many survivors face a lifetime of disability, including learning disabilities and visual and hearing problems. Although manual analysis of ultrasound images (US) is still prevalent, it is prone to errors due to its subjective component and complex variations in the shape and position of organs across patients. In this work, we introduce a conceptually simple convolutional neural network (CNN) trained for segmenting prenatal ultrasound images and classifying task for the purpose of preterm birth detection. Our method efficiently segments different types of cervixes in transvaginal ultrasound images while simultaneously predicting a preterm birth based on extracted image features without human oversight. We employed three popular network models: U-Net, Fully Convolutional Network, and Deeplabv3 for the cervix segmentation task. Based on the conducted results and model efficiency, we decided to extend U-Net by adding a parallel branch for classification task. The proposed model is trained and evaluated on a dataset consisting of 354 2D transvaginal ultrasound images and achieved a segmentation accuracy with a mean Jaccard coefficient index of 0.923 $\pm$ 0.081 and a classification sensitivity of 0.677 $\pm$ 0.042 with a 3.49\% false positive rate. Our method obtained better results in the prediction of preterm birth based on transvaginal ultrasound images compared to state-of-the-art methods.

\end{abstract}
\section{Introduction}

Preterm birth (PTB), defined as birth before 37 weeks of gestation, affects 5-18\% of pregnancies worldwide, which is equivalent to 15 million preterm neonates each year \cite{C1}. Despite major advances in perinatal care, preterm birth still accounts for 75\% of neonatal deaths and over 50\% of neurological handicap in children. Prediction and early detection of women at high risk of PTB are crucial as it allows timely intervention. Current diagnostic methods involve the collection of maternal characteristics and transvaginal ultrasound imaging conducted in the first and second trimester of pregnancy. Analysis of the ultrasound data is based on visual inspection of images by a gynaecologist, sometimes supported by hand-designed image features such as cervical length. Apart from that, it can be difficult, even for an expert, to identify relevant structures within the image. Due to the complexity of this process and its subjective component, approximately 30\% of spontaneous preterm deliveries are not correctly predicted. We gave an overview of the current state-of-art models for semantic segmentation, especially applied in the recent research on medical image segmentation \cite{C1a}. We introduced the major, popular network structures used for image segmentation: Fully Convolutional Network, U-Net and Deeplabv3 for cervix segmentation task. We evaluated all three proposed models against cervix segmentation task, compared its results and decided to proceed with U-Net as the one obtained best segmentation results. Recently, several works showed how quantitative methods based on U-Net-like architecture can successfully be used to perform a joint segmentation and classification task \cite{C6}, \cite{C7}. However, none of those methods are dedicated to preterm birth classification tasks.

In this paper, we address the problem of spontaneous preterm birth prediction. We introduce a novel approach in the context of preterm birth, that is inspired by linking classification and segmentation together.
As a contribution, we propose a deep-learning preterm prediction and cervix segmentation framework for ultrasound images. We show that extracted cervix features in the U-Net contracting path contains valid information for preterm classification task. To the best of our knowledge, this is the first one-stage work which employs deep convolutional neural networks (CNNs) for detecting preterm birth using transvaginal ultrasounds images.

Our method introduces objectively the obtained result, in contrast to the gynaecologist's manual approach, based solely on his knowledge. Unlike U-Net, our network is capable of detecting a preterm pregnancy and segmenting the cervix in an US image simultaneously. The preterm \textit{vs.} control type classification is implemented by feeding a part of the features in U-Net to a sequence of fully-connected layers followed by a sigmoid layer.


\section{Methods}

In this section, we first discuss dataset and preprocessing used in training and testing. Secondly, we present multi-task loss function for segmentation and classification task. Then, we propose an extension for one of neural network architecture for simultaneous segmentation and classification preterm \textit{vs.} control. Finally, we present Grad-CAM as visualization to increase understanding of the classification spontaneous preterm birth by a convolutional neural network.

\begin{figure}[ht!]
\begin{minipage}[b]{.5\linewidth}
  \centering
  \centerline{\includegraphics[width=4.2cm]{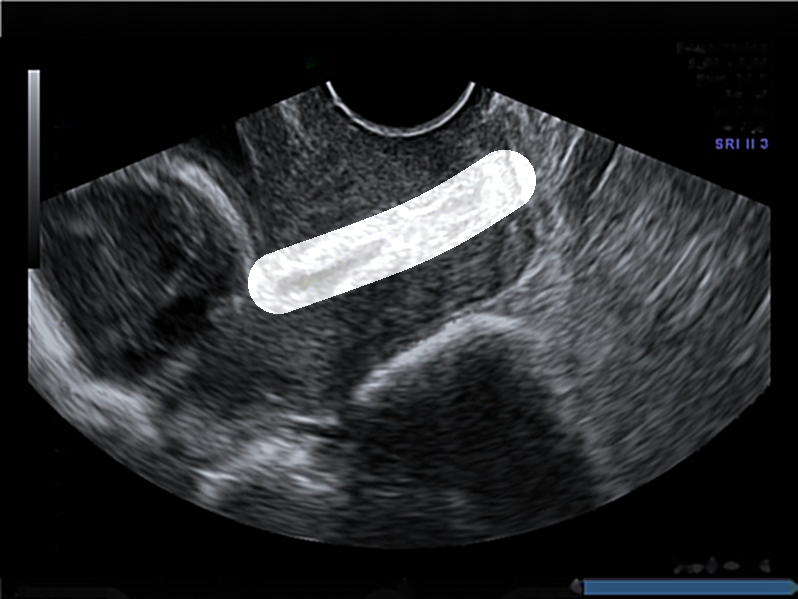}}
\end{minipage}
\hfill
\begin{minipage}[b]{0.49\linewidth}
  \centering
  \centerline{\includegraphics[width=4.2cm]{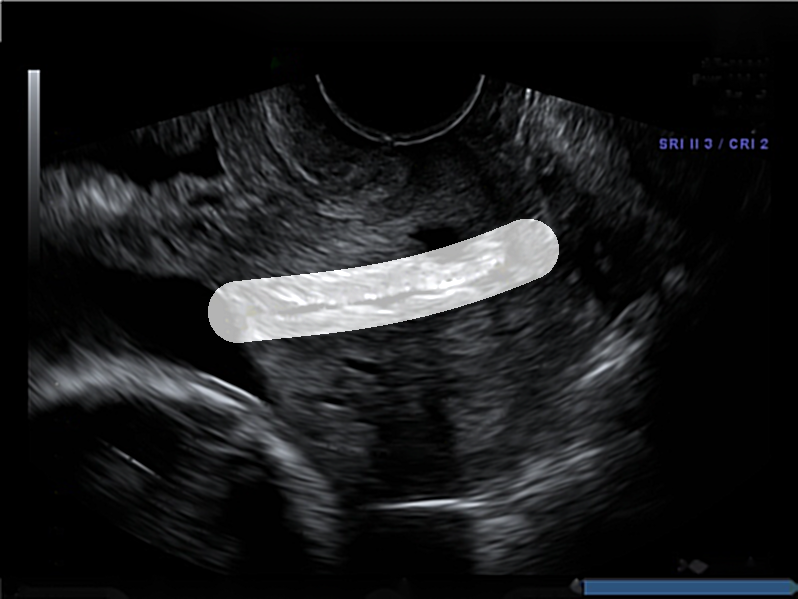}}
\end{minipage}
\caption{Example of our 2D transvaginal ultrasound images used as a part of dataset. Input images and their masks showing the location and shape of the cervix}
\label{fig:dataset}
\end{figure}

\textbf{Dataset:} The original dataset consists of 354 two-dimensional transvaginal cervical ultrasound images. Data were collected from female volunteers during check-ups conducted in the first and second trimesters of pregnancy. Data includes 319 images of the pregnancy control group and 35 images of premature deliveries which reflects the statistical occurrence of this phenomenon in reality. Note that to get information about image labels, we waited for the pregnancy result of each patient.
The data was provided by two different clinics: King's College London and Medical University of Warsaw. All images have been annotated by several independent gynecologist experts via our own annotation tool. 
Sample images from the data set are depicted in Fig. \ref{fig:dataset}.

The annotations received contained the study identifier and four control points which allowed for the drawing of cubic B\'ezier curves that outlined the shape and position of the cervix in the image. B\'ezier curves were used to create masks for cervical segmentation.

\textbf{Preprocessing:} We first removed the manually embedded annotations in the form of yellow and green crosses placed by gynecologists during routine ultrasound examinations. For this purpose, we used the inpainting algorithm proposed by Telea \cite{C8}, which was used in a similar case by \cite{C9}. This allowed our algorithm to learn from the images rather than learning manually embedded annotations. Note, these annotations were unrelated to the task we were doing. In order to improve model generalization we divided the dataset into three parts: training, test, and validation. Then we did a 50:50 data augmentation between the preterm class and control on each subset to avoid heavily focusing on the majority class by classification algorithm. We applied various types of data augmentation techniques such as cropping, random rotation in the range of -15 to +15 degrees and adding contrast, brightness or noise. As a result the set was expand to 6354 images.
On the dataset, based on created annotations for each examination by several gynecological experts, we created segmentation masks, which were used to train the neural network in the segmentation task.

\textbf{Loss function:} We define a multi-task total loss as 
\begin{equation}
    \mathcal{L} = \mathcal{L}_{Seg} + \mathcal{L}_{Cls}
\end{equation}
where $\mathcal{L}_{Seg}$ and $\mathcal{L}_{Cls}$ are the binary cross-entropy loss (BCE) functions for the segmentation and classification tasks, respectively. Where $y$ is the ground truth label, $\hat{y}$ is the prediction, BCE is defined as follows:
\begin{equation}
    \textit{BCE}(y, \hat{y}) = y \cdot log\hat{y} + (1-y) \cdot log(1-\hat{y})
\end{equation}
In order to improve the segmentation accuracy, we use the negative value of Dice coefficient (called Dice loss) as well as the criterion to optimize the model parameters. The Dice score coefficient (DSC) is a measure of overlap widely used to assess segmentation performance. Proposed in Milletari et al. \cite{C9a} as a loss function, it can be expressed as
\begin{equation}
    \mathcal{L}_{Dice}=1-\frac{2 \sum y
    \hat{y}}{\sum y^{2}+\sum \hat{y}^{2}+\epsilon}
\end{equation}
The $\epsilon$ term is used here to ensure the loss function stability by avoiding the numerical issue of dividing by 0.

\noindent Total segmentation loss is defined as:
\begin{equation}
    \mathcal{L}_{Seg} = \alpha \cdot \textit{BCE} + (1 - \alpha) \cdot \mathcal{L}_{Dice}
    \label{eq:seg_loss}
\end{equation}
The BCE loss itself tries to match the background and foreground pixels in the prediction to the ground truth masks. BCE loss does not emphasize on keeping the object together while Dice loss considers the entire object by computing the overlap between predicted and ground truth objects. Thus combining both BCE loss and the Dice loss will give us the advantages of both. For the classification task, to account for the class imbalance between the \textit{preterm} and \textit{term} classes, we introduced a stronger weight in the loss function for the under-represented class (\textit{preterm}).
\begin{equation}
    \mathcal{L}_{Cls} = \beta \cdot \textit{BCE}_{w} 
\end{equation}
In the experiment, $\alpha$ is set to 0.5, and $\beta$ to 0.8.

\textbf{Training:} In the experiment, we selected 3812 images (60\%) for training, 1271 images (20\%) for validating and 1271 images (20\%) for testing. Note, after augmentation, all images of a given patient were in the same set: either training, test or validation. We resized all images into 256 px$\times$256 px. We trained a network on a machine with AMD FX-8320 @ 3.5GHz CPU and NVIDIA TITAN X 12GB GPU and implemented our models using the PyTorch library with CUDA support. 

We trained all networks for 500 epochs with a batch size of 4. We chose Adam optimizer with the default parameters ($\beta_{1}$ = 0.9 and $\beta_{2}$ = 0.999), learning rate of $10^{-4}$ and weight decay of $5\cdot10^{-4}$. The training took 15 hours.

\begin{figure}[ht!]
    \centering
    \includegraphics[width=12cm]{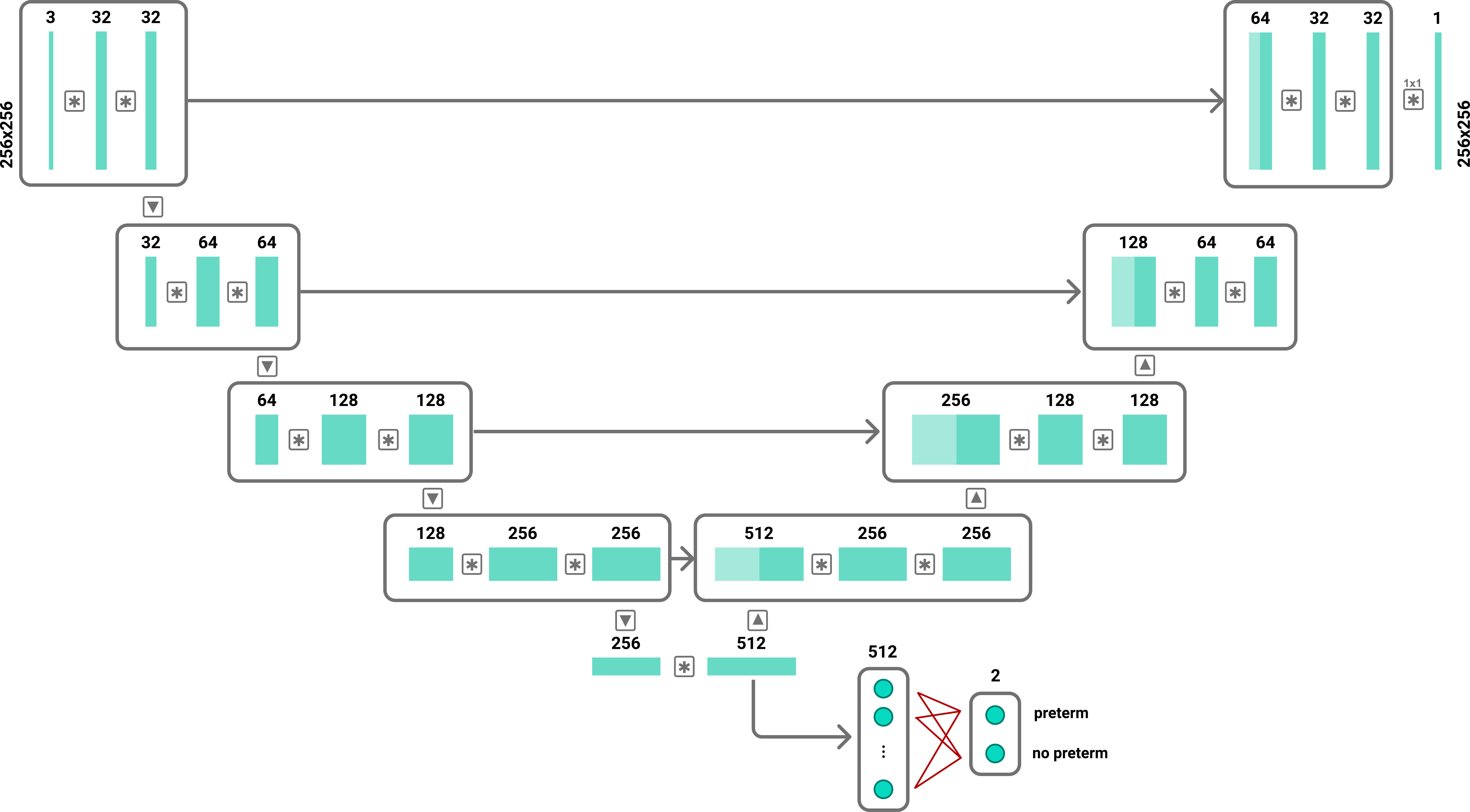}
    \caption{Neural network architecture for simultaneous segmentation and classification for end-to-end spontaneous birth preterm classification.}
    \label{fig:fig-arch}
\end{figure}

\textbf{Visualization:} In order to increase the understanding of the predictions of spontaneous preterm births performed by the convolutional neural network on ultrasound images, we have visualized them by using the gradient weighted class activation mapping (Grad-CAM) method. This method is used for visual interpretation of deep neural networks through a gradient-based location, highlighting the most interesting regions in the image for prediction. Grad-CAM uses gradient information provided to the last layer of the CNN to allow the viewer a greater understanding of the importance of each neuron's region of interest. By using Grad-CAM, it is possible to visualize which features in the cross section contribute to judging the class.
For this purpose, we used the original implementation of the work of Selvaraju et al. \cite{C12}.

\section{Results}

In this section, we evaluate three described neural network models and check how they perform in the task of cervix segmentation. Then, based on obtained results, the best performing model is selected, extended and evaluated for classification task.

\subsection{Image Segmentation}

We validated three models against the cervix segmentation task applied to transvaginal ultrasounds images: FCN, DeepLab and U-Net. Visualization of results for segmentation are presented on Fig. \ref{fig:seg-results}. As shown in Table \ref{tab:three_model_seg_results}, the best performing results were achieved by U-Net. U-Net is built upon the architecture of FCN. Besides the increased network depth to 19 layers, U-Net benefits from a superior design of skip connections between different stages of the network. The most important advancement of U-Net over FCN is the shortcut connections between layers of equal resolution in the analysis path to expansion path. This gives U-Net better performance because it has multiple upsampling layers along with more skip connections which make it more robust to scale variations as compared to FCN.

DeepLab, on the contrary, uses convolutions with upsampled filters - dilated convolutions - to achieve better control over the feature response resolution and to incorporate larger context without increasing the computational cost. However, the discontinuity between the dilated convolution kernel leads to the omission of some pixels, which may lead to the neglect of the continuity information on the image \cite{C10}. This means that whereas increasing dilation factors is important in terms of resolution and context, it can be detrimental to small objects, especially on noisy ultrasound images. 

We chose the U-Net based network for the classification task for several reasons. First, we obtained the best segmentation result qualitatively. The U-Net network best predicts edges that help us classify images based on the shape of the cervix. In addition, the U-net network contains the least parameters among the other networks tested, without requiring a large number of images and length of training to get a good performance.

\begin{table}[ht!]
\caption{Mean IoU and std scores of U-net, Fully Convolutional Network and DeepLab Net. The three models were trained for 500 epochs and use the same loss function (\ref{eq:seg_loss})}
\centering
\begin{tabular}{m{2.2cm}ccc}
\hline
& \textbf{U-Net} & \textbf{FCN} & \textbf{DeepLab}  \\ 
\hline
\midrule
\multirow{2}{*}{Train}
 & \textbf{0.975} & 0.848 & 0.813 \\
 & (0.041) & (0.062) & (0.056) \\
\midrule
\multirow{2}{*}{Val}
 & \textbf{0.921} & 0.770 & 0.791 \\
 & (0.042) & (0.023) & (0.086)  \\
 \midrule
 \multirow{2}{*}{Test}
 & \textbf{0.913} & 0.755 & 0.756 \\
 & (0.041) & (0.042) & (0.087) \\
\hline
\end{tabular}
\label{tab:three_model_seg_results}
\end{table}%
 
\newcommand\columnWidth{0.2\textwidth}
\newcommand\itemWidth{2.1cm}

\setlength\tabcolsep{2pt}%
\begin{figure}[ht!]
   \centering
\begin{tabular}{ccccc}
\includegraphics[width=\itemWidth]{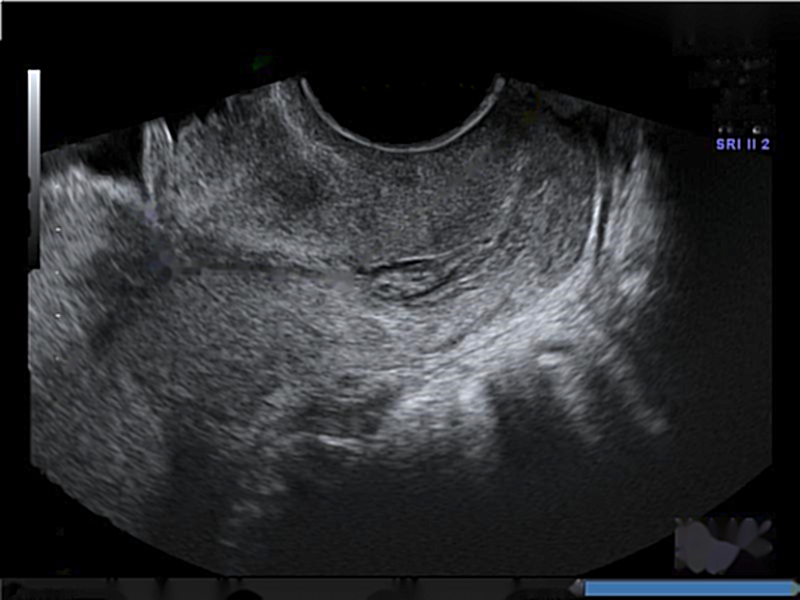}&
\includegraphics[width=\itemWidth]{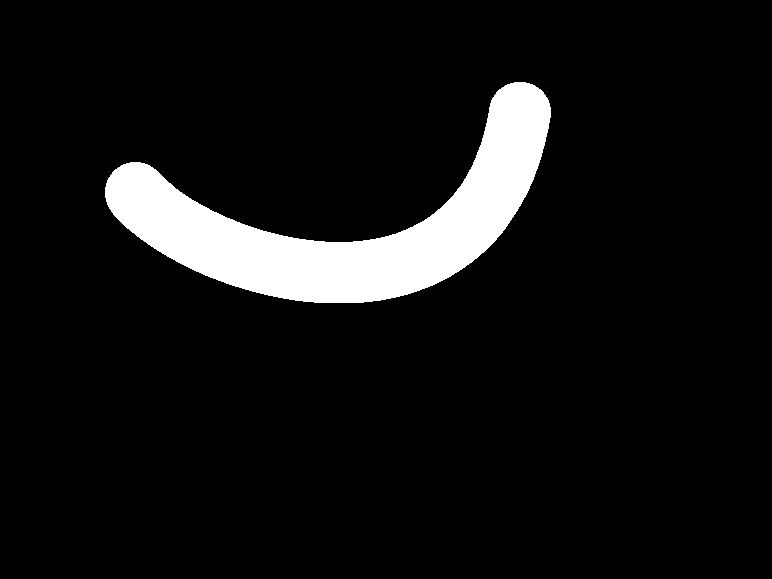}&
\includegraphics[width=\itemWidth]{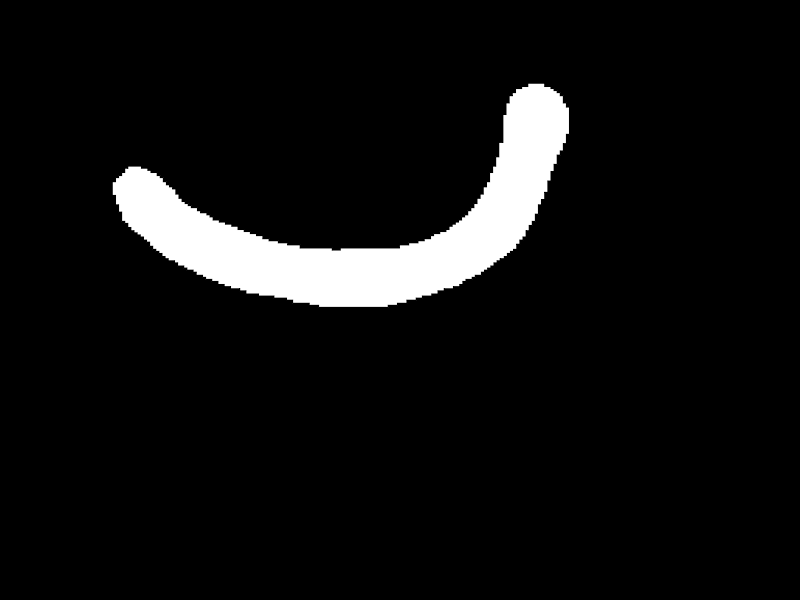}&
\includegraphics[width=\itemWidth]{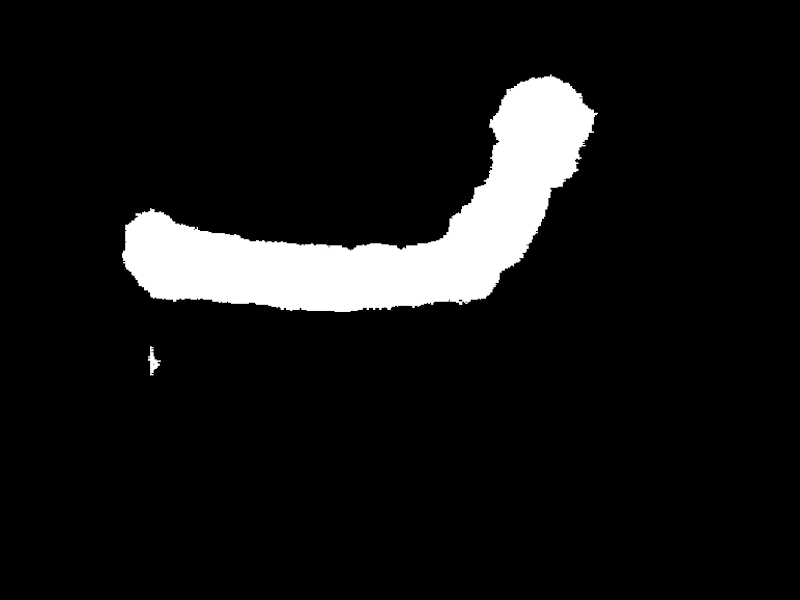}&
\includegraphics[width=\itemWidth]{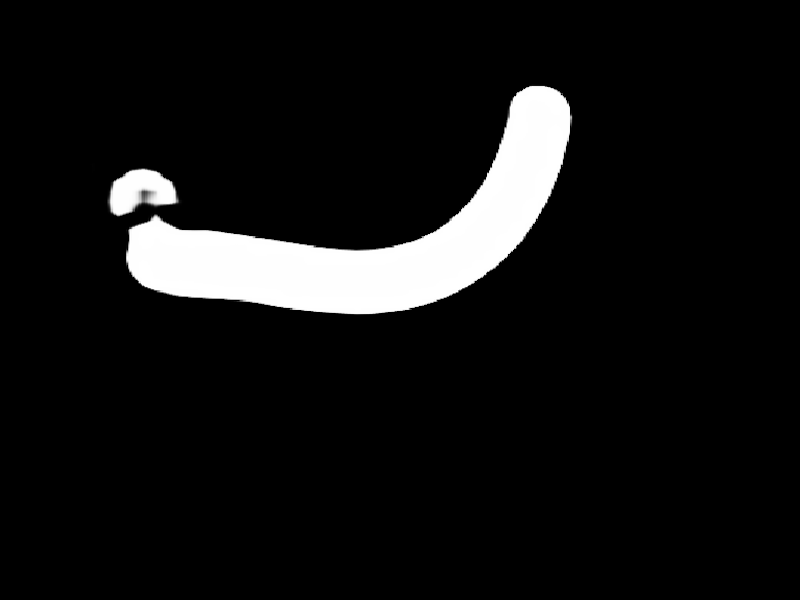}\\
\includegraphics[width=\itemWidth]{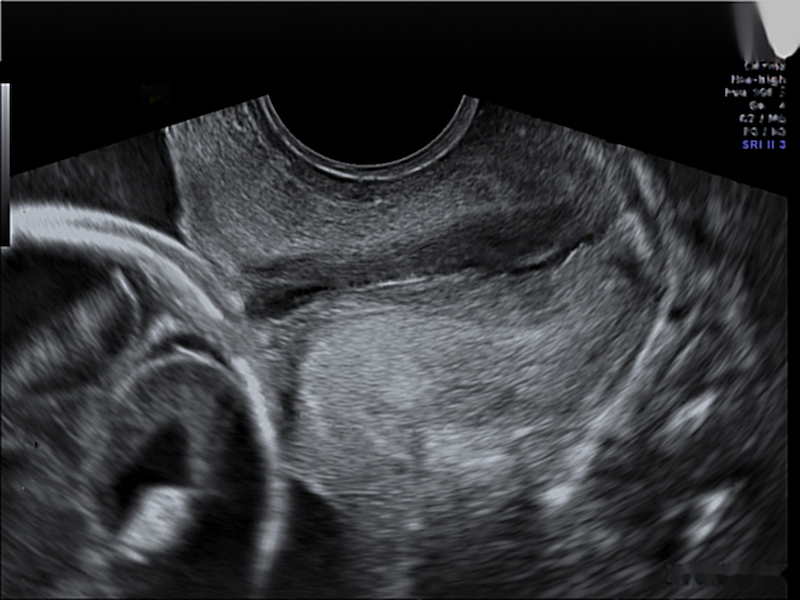}&
\includegraphics[width=\itemWidth]{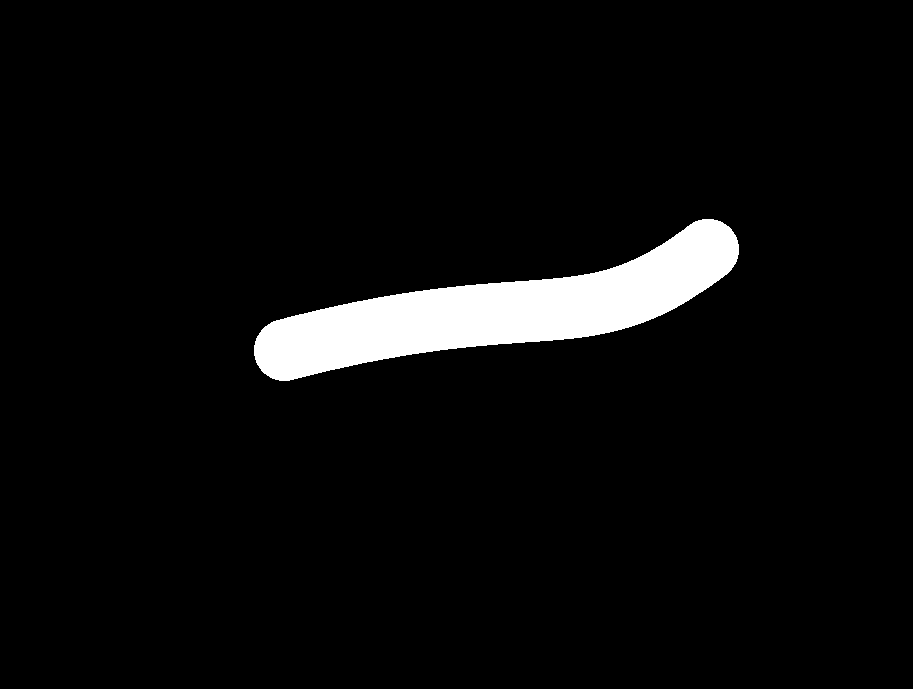}&
\includegraphics[width=\itemWidth]{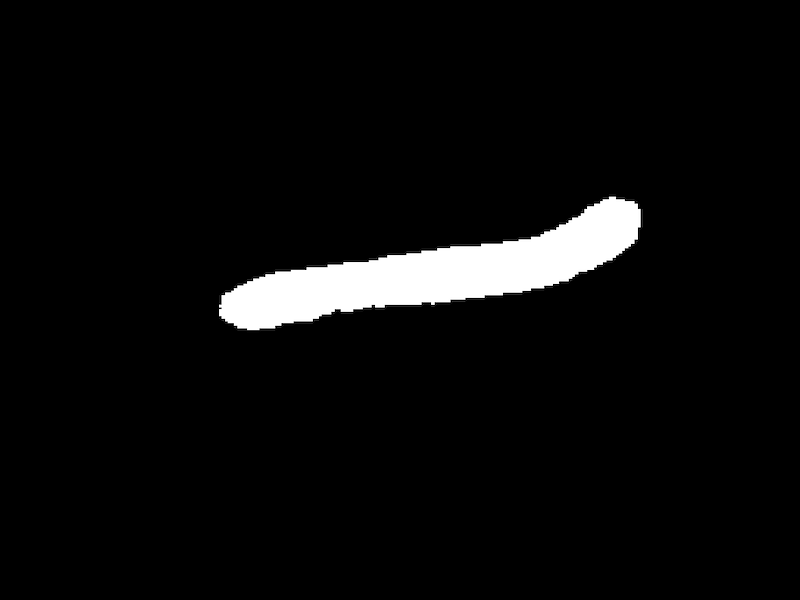}&
\includegraphics[width=\itemWidth]{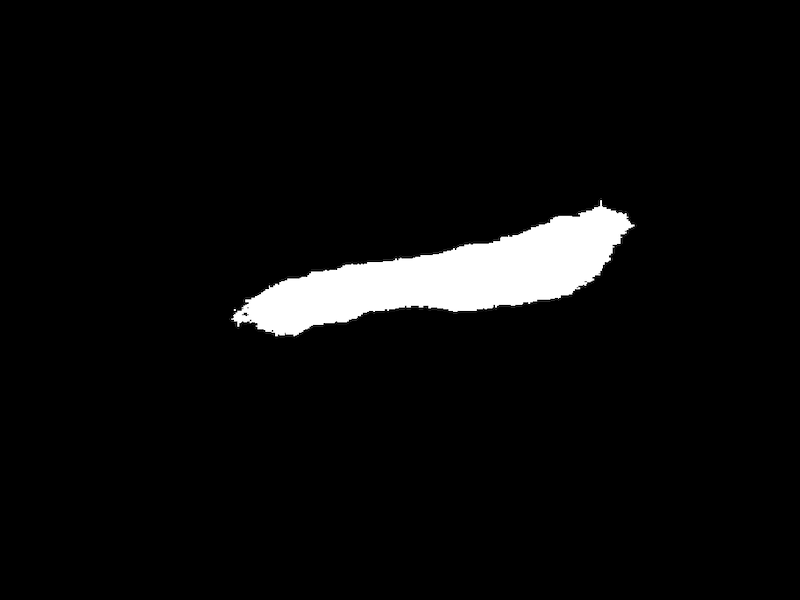}&
\includegraphics[width=\itemWidth]{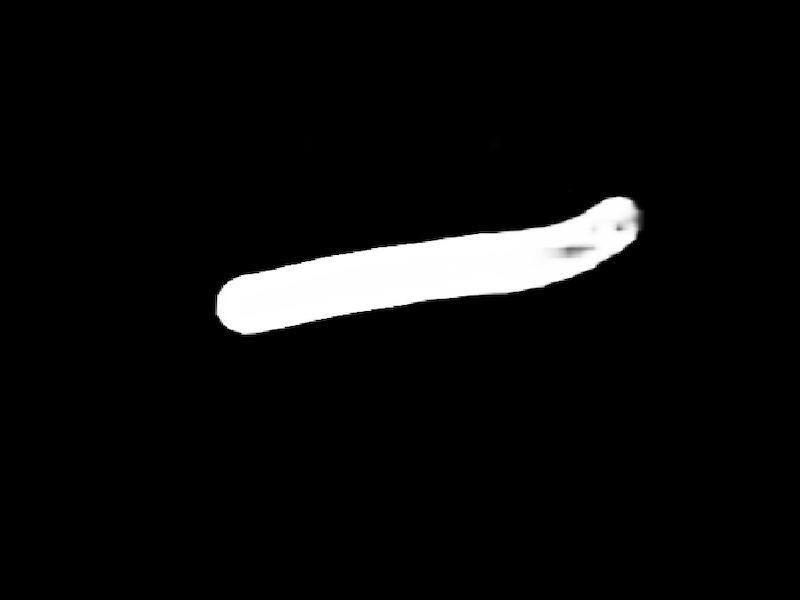}\\
\includegraphics[width=\itemWidth]{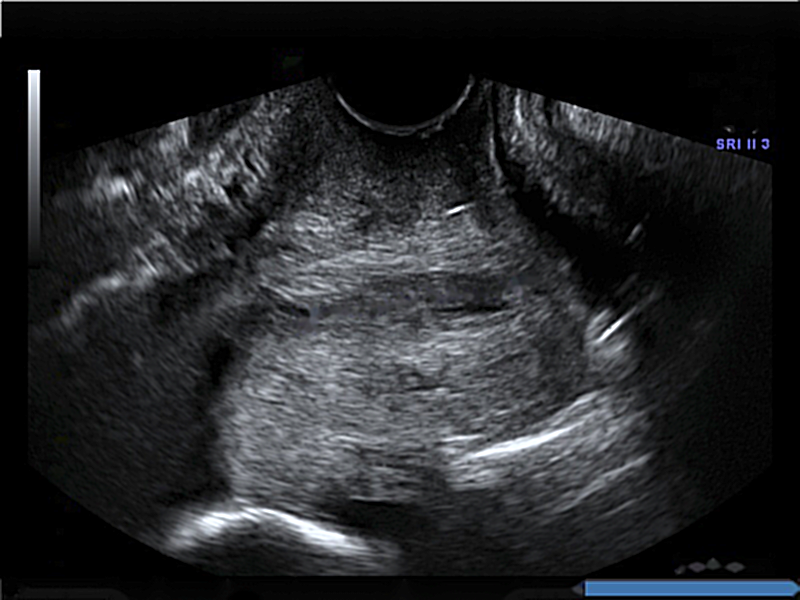}&
\includegraphics[width=\itemWidth]{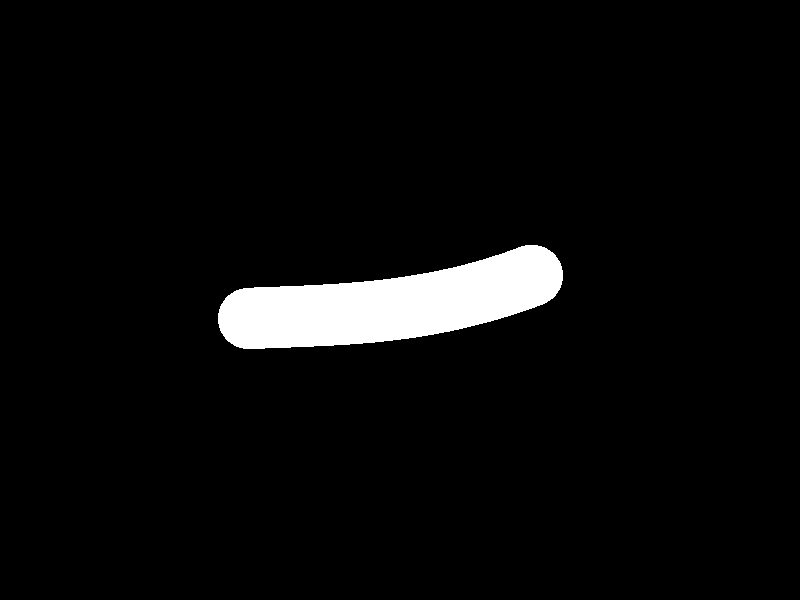}&
\includegraphics[width=\itemWidth]{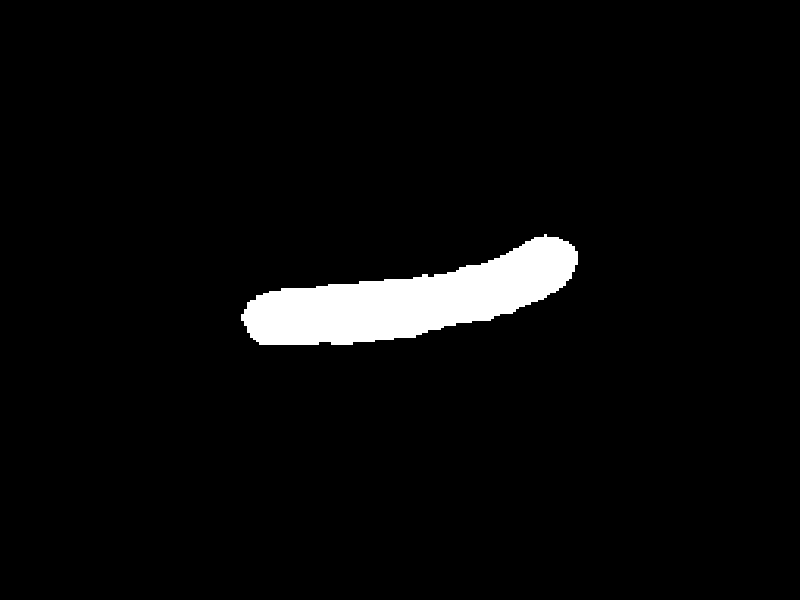}&
\includegraphics[width=\itemWidth]{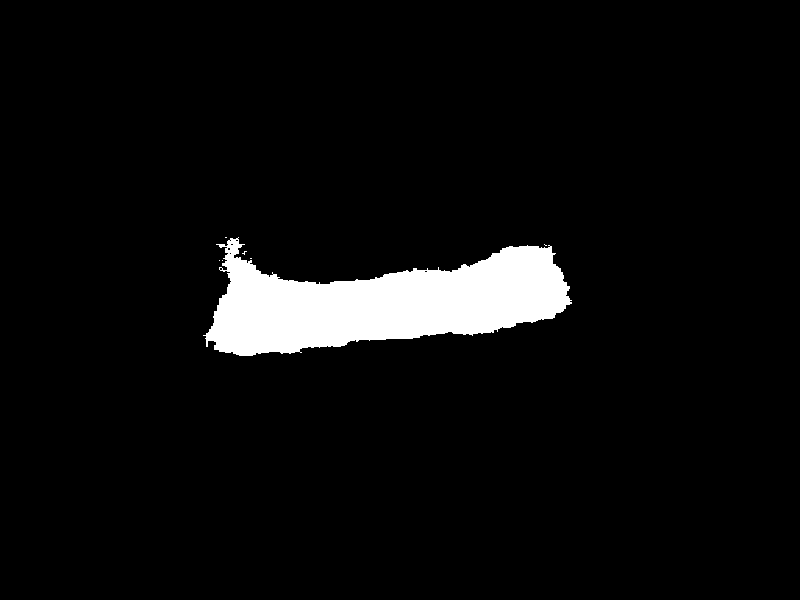}&
\includegraphics[width=\itemWidth]{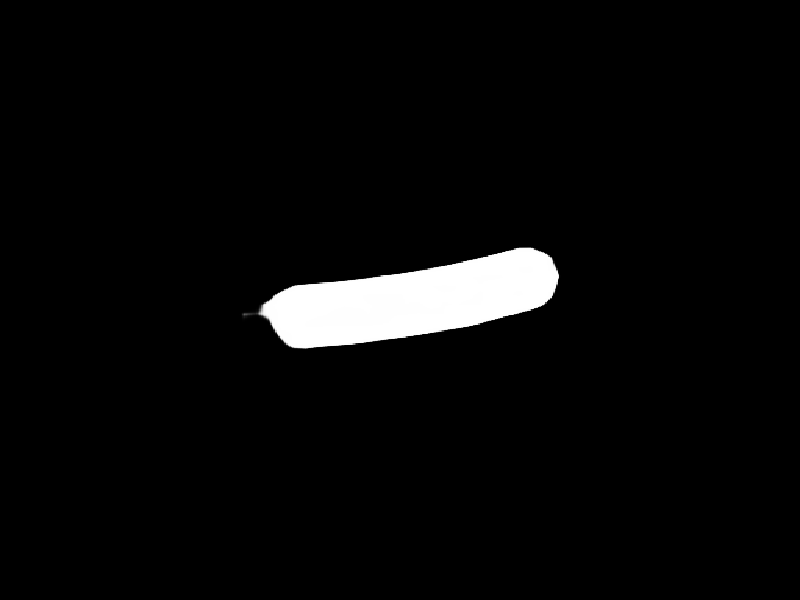}\\
(a)&(b)&(c)&(d)&(e)\\
\end{tabular}

    \caption{Visualization of results for segmentation: a) Input image b) Ground truth mask c) U-Net d) FCN e) Deeplabv3}
    \label{fig:seg-results}
\end{figure}

\begin{figure*}[ht!]
  \centering
  \centerline{\includegraphics[width=0.95\textwidth]{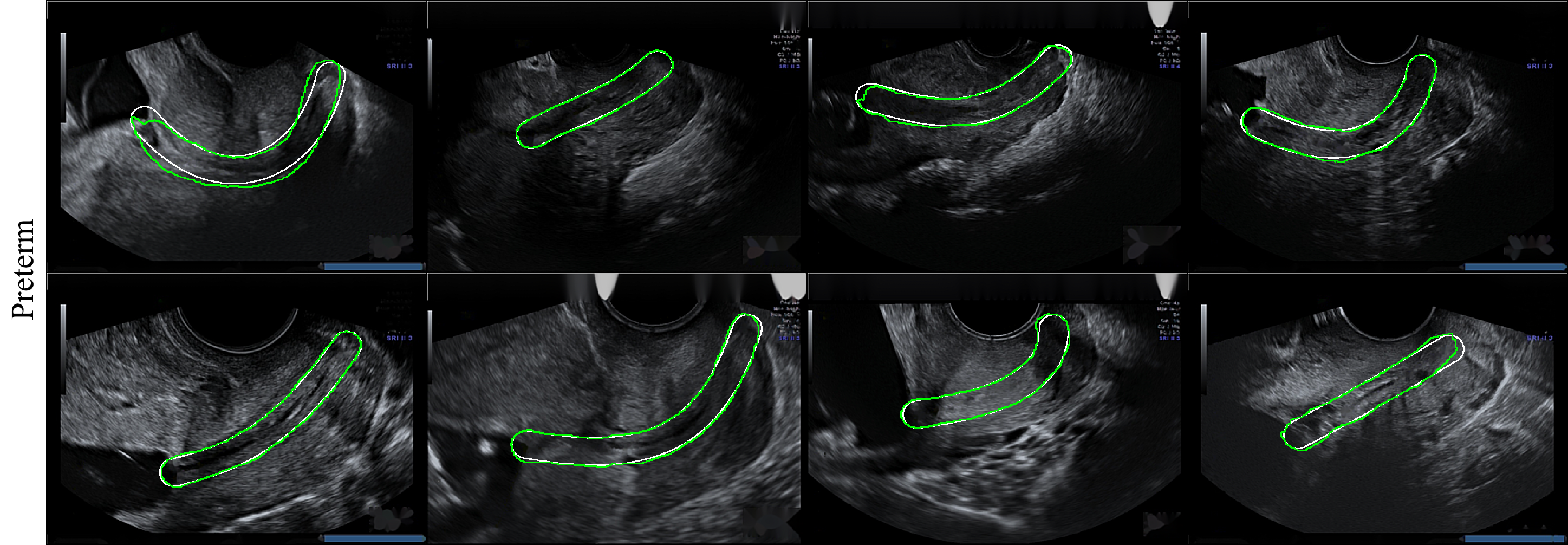}}
\end{figure*}

\begin{figure*}[ht!]
  \centering
  \centerline{\includegraphics[width=0.95\textwidth]{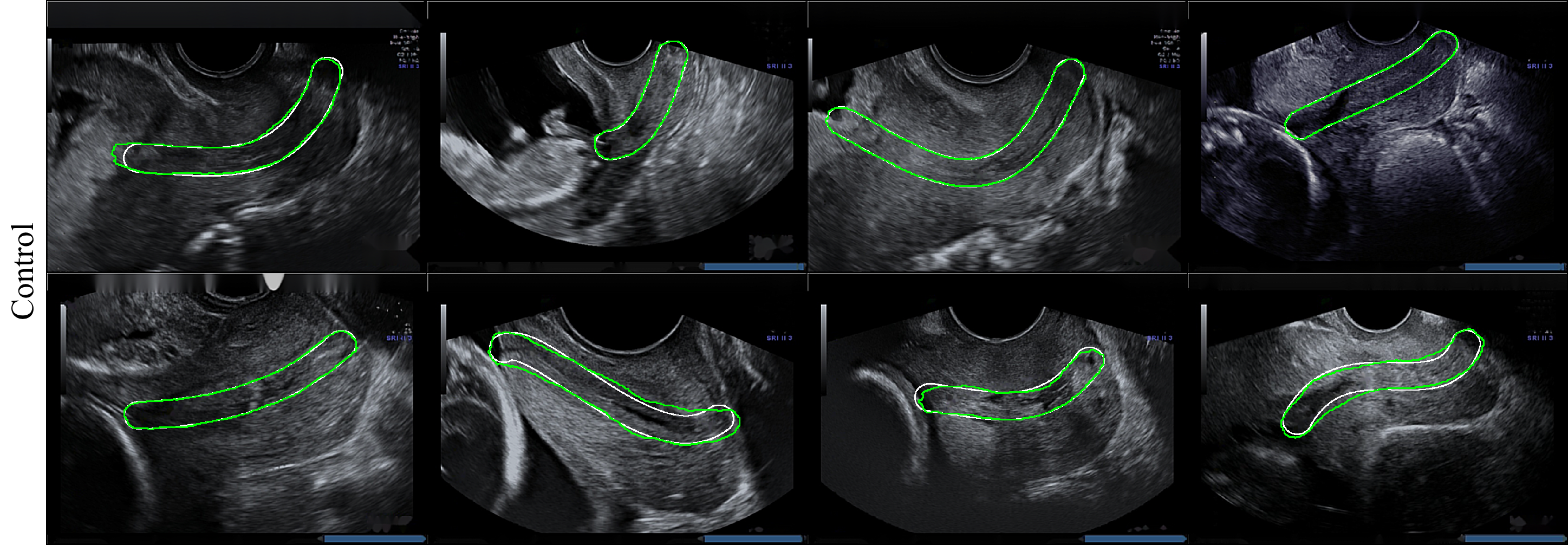}}
\caption{Results of cervical segmentation on transvaginal ultrasound images using the segmentation part of the neural network architecture. White represents ground truth, while green represents our predictions. We obtained an average Jaccard index of 92.3, with a standard deviation of 0.081}
\label{fig:results}
\end{figure*}

\subsection{Classification}

We achieved good segmentation results due to the easy structure of the cervix, which is relatively large in the ultrasound image. In Fig. \ref{fig:results} we show examples of retrieved cervix segmentation masks for each of the classes. From these examples, it can be seen that our proposed method was able to localise cervixes which are subject to variability in scale and appearance. Table \ref{tab:class_results} shows the quantitative performance of the proposed method in terms of mean $\pm$ std of IoU coefficient, precision, recall and AUC.

\renewcommand{\arraystretch}{1.1}

\begin{table}[ht!]
\caption{Segmentation (mIoU) and classification results (Recall, Precision, AUC) (mean and std) for train, validation and test dataset}
\centering
\begin{tabular}{m{1.9cm}cccc}
\hline
& \textbf{IoU} & \textbf{Recall} & \textbf{Precision} & \textbf{AUC} \\ 
\hline
\midrule
\multirow{2}{*}{Train}
 & 0.972 & 0.781 & 0.772 & 0.782 \\
 & (0.047) & (0.092) & (0.096) & (0.108) \\
\midrule
\multirow{2}{*}{Val}
 & 0.925 & 0.698 & 0.698 & 0.757 \\
 & (0.072) & (0.023) & (0.086) & (0.156)  \\
 \midrule
 \multirow{2}{*}{Test}
 & 0.923 & 0.677 & 0.683 & 0.723 \\
 & (0.081) & (0.042) & (0.087) & (0.134)  \\
\hline
\end{tabular}
\label{tab:class_results}
\end{table}%

Our method obtained better results according to Intersection over Union (IoU) in segmentation, recall and precision for the classification task than \cite{C11}. Table \ref{tab:class_baseline} presents a comparison of our results with the baseline - IoU for segmentation and recall, precision, AUC for the classification task. The confusion matrix for the preterm birth classification is shown in Table \ref{tab:confusion_matrix}. We obtained specificity of 0.951 and sensitivity of 0.68 with a 3.49\% false positive rate.
\newline

\begin{table}[ht!] 
\caption{Comparison of the results obtained between state-of-the-art \cite{C11} and ours}
\centering
\begin{tabular}{m{2.7cm}cccc}
\hline
& \textbf{IoU} & \textbf{Recall} & \textbf{Precision} & \textbf{AUC} \\
\hline
\midrule
\multirow{1}{*}{W\l odarczyk et al.}
 & 0.91 & 0.596 & 0.659 & 0.78 \\
\midrule
 \multirow{1}{*}{Our}
 & \textbf{0.923} & \textbf{0.677} & \textbf{0.683} & 0.723 \\
\hline
\end{tabular}
\label{tab:class_baseline}
\end{table}%

\begin{table}[ht!]
\caption{The confusion matrix for the preterm birth classification task.}
\centering
\noindent
\renewcommand\arraystretch{1.1}
\begin{tabular}{cc|cc}
\multicolumn{2}{c}{}
            &   \multicolumn{2}{c}{Predicted} \\
    &       &   Control &   Preterm             \\ 
    \cline{2-4}
\multirow{2}{*}{\rotatebox[origin=c]{90}{Actual}}
    & Control   & 1105   & 40                 \\
    & Preterm    & 41    & 85                \\ 
    \cline{2-4}
    \end{tabular}
\label{tab:confusion_matrix}
\end{table}

\subsection{Grad-CAM}

\begin{figure}[ht!]
   \centering
\begin{tabular}{c}
\includegraphics[width=0.7\textwidth]{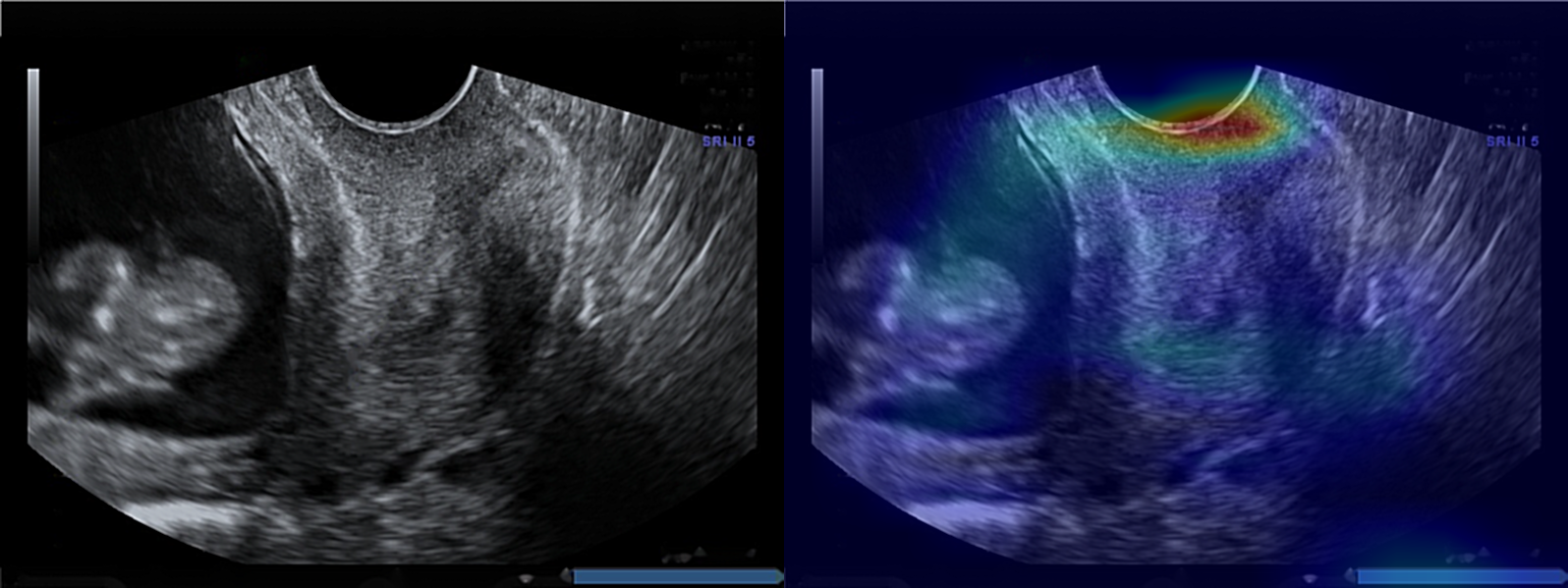}\\
\includegraphics[width=0.7\textwidth]{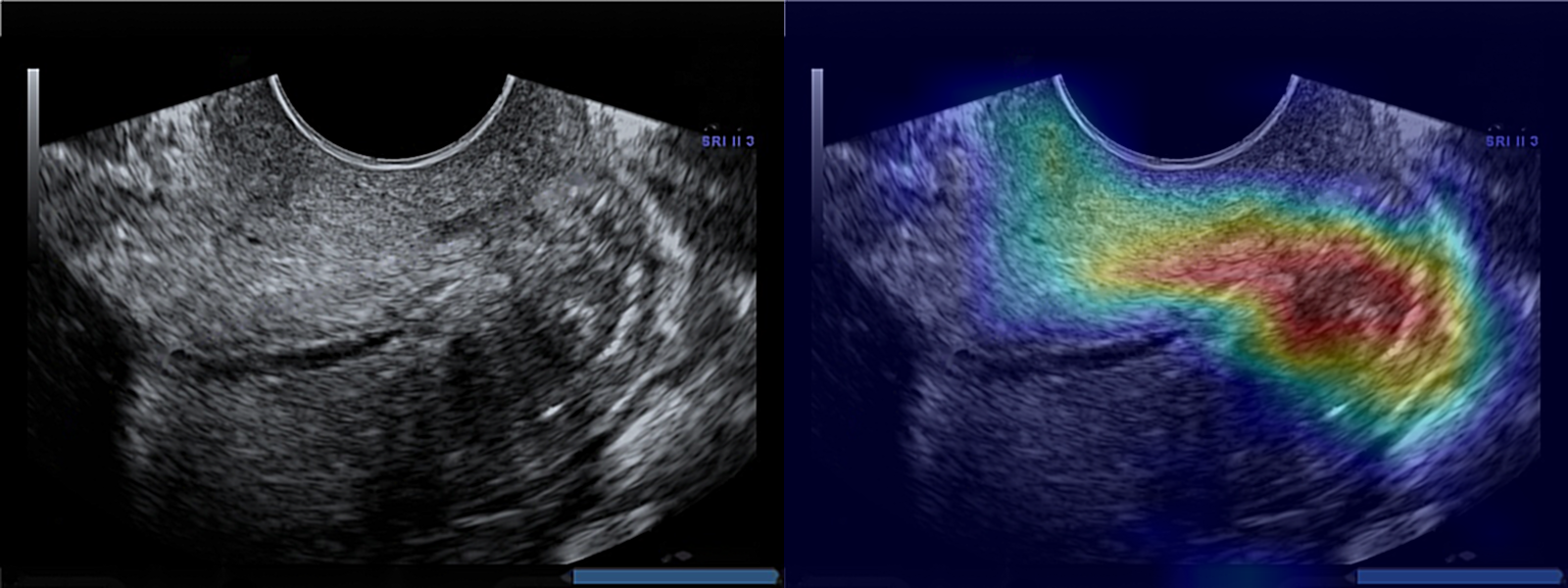} 
\end{tabular}
    \caption{Visual interpretation of CNNs through a gradient-based location, highlighting the most interesting regions in the image for prediction control (top) and preterm (bottom) class. The decision of the classifier is made based on the central parts of the ultrasound image which is similar to the analysis done manually by the physicians}
    \label{fig:grad_cam}
\end{figure}

It was found by Baños et al. \cite{C13} that information extracted from the region along the length of the anterior cervical stroma is relevant to control vs preterm classification. In addition, Pachtman et al. \cite{C14} proved that the region along the length of the anterior cervical stroma and the analysis of its relative organization of cervical collagen fibers may have the capacity to identify preterm birth.
To the contrary, Grad-CAM shows the preterm class is classified based on the lower segment of the cervix, close to the ectocervix. We can note that during classification, the neural network focuses on the heterogeneity of the density of tissues around the cervix. For the control class, our model is focusing on the top part of the largest homogeneous region in middle part of anterior cervical lip, which is the part of the cervix closest to the transducer. Examples of Grad-CAM results can be seen in Figure \ref{fig:grad_cam} for the preterm and control class. 

\renewcommand\itemWidth{4.2cm}
\setlength\tabcolsep{2pt}

\section{Conclusions}
The goal of this paper was to evaluate whether convolutional neural networks can be successfully applied to preterm birth prediction tasks, especially considering analysis of transvaginal ultrasound images. Conducted experiments show that a convolutional model outperforms common approaches based on manually engineered features \cite{C15} and achieves a similar level as other feature learning methods.

The proposed method achieves a segmentation accuracy with a mean Jaccard coefficient index of 0.923 $\pm$ 0.081 and a classification sensitivity of 0.68 $\pm$ 0.042. To the best of our knowledge, this is the best result of a segmentation and classification method for spontaneous preterm birth prediction using transvaginal ultrasound images. The results presented in this paper show that methods based on deep neural networks can provide automatic, quantitative analysis of ultrasound images. This, in turn, can lead to significant time savings and increase the efficiency of current diagnostic methods without losing its precision.

Our proposal, an objective method for detection preterm birth, may help to identify patients at risk for sPTB before any changes reflected by CL measurement. Although our results are encouraging, a prospective, longitudinal study is necessary to validate this technique, allow us to avoid selection bias at enrollment and control for potential confounding variables related to treatment nonuniformity. In the future, we want to reduce the size of false negative error by increasing the size of the dataset and employing UNet-3D for deeper analyses of ultrasound images. 

%
%
%
%

\end{document}